\documentclass{optica-article}

\journal{opticajournal} 

\articletype{Research Article}

\usepackage{lineno}

\begin{document}

\title{Design of a monolithic source of photon pairs comprising a semiconductor laser and a Bragg reflection waveguide}

\author{Thomas Tenzler,\authormark{1,*} Jan-Philipp Koester,\authormark{1} Hans Wenzel,\authormark{1} Thorsten Passow,\authormark{2} Quankui Yang,\authormark{2} Marko Haertelt,\authormark{2} and Andrea Knigge\authormark{1}}

\address{\authormark{1}Ferdinand-Braun-Institut (FBH), Berlin, Germany\\
\authormark{2}Fraunhofer-Institut für Angewandte Festkörperphysik (IAF), Freiburg, Germany}

\email{\authormark{*}thomas.tenzler@fbh-berlin.de} 

\begin{abstract*} 
We propose a monolithic, electrically driven source of photon pairs based on a non-linear AlGaAs Bragg reflection waveguide and a laser structure stacked on top. By introducing lateral tapers, the fundamental mode of the lasing waveguide is vertically coupled into a higher order mode of the Bragg reflection waveguide (Bragg mode) such that photon pairs can be generated through a type-II spontaneous parametric down conversion process. According to numerical simulations, a coupling efficiency of 28\% is achieved between both modes. Phase matching the Bragg mode with two fundamental modes at 1550 nm results in a photon pair rate of $1.7\cdot10^{8}$ pairs/s for a 2 mm long device assuming 1 mW of power in the Bragg mode. Since the Bragg reflection waveguide does not require doping for this vertically coupled structure, free-carrier absorption losses and parasitic luminescence are avoided. 
\end{abstract*}

\section{Introduction}
Sources of photon pairs are essential for quantum-based encryption schemes such as quantum key distribution \cite{o2009photonic}. These correlated pairs offer a physically secure method of encrypting messages. Furthermore, they benefit from already established infrastructure when operating at wavelengths that are generally used for telecommunication. In the context of miniaturization, the electrically driven Bragg reflection waveguide (BRW) laser based on AlGaAs represents a promising candidate for this purpose. The device generates lasing light around 775 nm, which is then converted into correlated photon pairs around 1550 nm through a process known as spontaneous parametric down-conversion (SPDC) \cite{tenzler2025theoretical, tenzler2025tunable, boitier2014electrically, leger2025deterministically}. This provides a compact and easy to use source of correlated photon pairs which can subsequently be utilized in encryption schemes involving entangled photon pairs. However, the realization of such devices requires the incorporation of an active region, as well as an appropriate doping profile into the BRW. The resulting losses from free carrier absorption will reduce the photon pair output. The carrier injection leads to parasitc luminescence and a reduction of the Coincidence-to-Accidential Ratio (CAR) \cite{leger2025deterministically}. Phase matching required for the SPDC process necessitates compromises to obtain both a high laser efficiency and a high nonlinear efficiency.

We propose an alternative approach where the waveguide is spatially separated into an active section and passive section with each being optimized for its specific purpose, namely the generation of lasing light or SPDC. Then, in order to integrate the lasing and SPDC functionalities of the waveguide, lateral tapers are introduced that enable efficient vertical coupling between modes of the active and passive waveguides \cite{gerini2021electrically}. In addition to waveguide coupling, this type of integration by lateral tapers has also been demonstrated as a spot-size converter for fiber-coupling diode lasers \cite{kawano2002design}.

In the context of nonlinear processes such a structure with active and passive sections has been previously theoretically investigated as a candidate for an optical parametric oscillator based on AlGaAs. There, the nonlinear waveguide consists of a cladded GaAs core that is phase matched at around 975 nm \cite{gerini2021electrically}. As frequency degenerate photon pair generation around 1550 nm requires lasing at 775 nm, at which GaAs is not transparent, we propose a different nonlinear waveguide based on an asymmetric BRW. This waveguide enables phase matching between a TE-polarized high-order vertical waveguide mode and two perpendicularly polarized fundamental modes at 1550 nm. Both the InGaP n- and GaAs p-contacts of the laser waveguide are accessible from the top of the device, analogous to the technology reported in Ref. \cite{koester2025design}. 

Furthermore, neither an active region nor a doping profile have to be introduced into the bottom BRW, which mitigates losses from the absorption of photon pairs by free carriers. This is not the case in the recently demonstrated GaAs-based PIC platform shown in Ref.~\cite{koester2024gaas}. There, active and passive waveguides were realized through selective quantum well removal and two-step epitaxy. In contrast, the stacked waveguide approach presented here relies on a single growth step, as the lasing and photon pair generation functionalities are integrated via vertical coupling through lateral tapers.

The paper is structured as follows. In section \ref{Section_Design} details are provided about the design and the modes of the vertical structure. In section \ref{Section_Simulation} the operating principle of the tapers are shown and their coupling efficiency is calculated. In section \ref{Section_Results} the nonlinear properties of the passive BRW are discussed. In section \ref{Section_Conclusion} the work is concluded. 
\section{Design of the laser structure}
\begin{figure}
	\centering
	\includegraphics[scale=0.5]{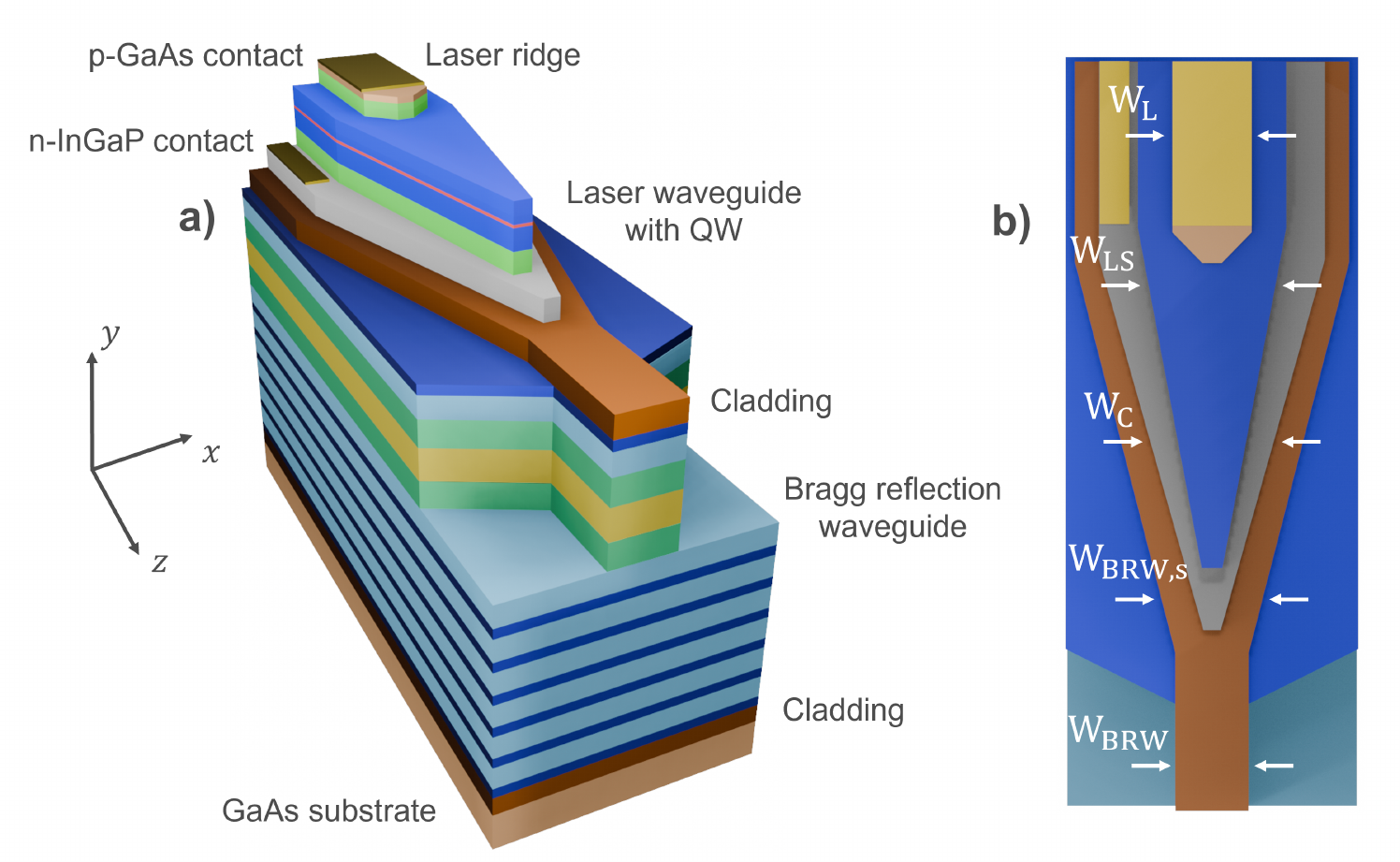}
	\caption{a) 3D-schematic of the tapered stacked waveguide laser based on AlGaAs. b) Top-view of the 3D-schematic shown in Fig. \ref{Schematic_Taper}a. The relevant widths within the waveguides and tapers are shown by $W$.}
	\label{Schematic_Taper}
\end{figure}
The stacked waveguide approach enables the separate optimization of an active section and a passive section in a waveguide. As shown in Fig. \ref{Schematic_Taper}, the structure consists of a bottom BRW and a 775 nm laser structure stacked on top. Current injection is realized between the p-contact on top of the structure and the n-contact in between the laser waveguide and the BRW.  

The layer-stack of the undoped BRW consists of GaAs substrate, Al$_{0.85}$Ga$_{0.15}$As cladding layer, six pairs of Bragg reflector layers with alternating aluminum content Al$_{0.25}$Ga$_{0.75}$As/Al$_{0.61}$Ga$_{0.39}$As, two Al$_{0.25}$Ga$_{0.75}$As matching layers surrounding an Al$_{0.43}$Ga$_{0.57}$As core layer, followed by one single pair of Bragg reflectors Al$_{0.61}$Ga$_{0.39}$As/Al$_{0.25}$Ga$_{0.75}$ and an Al$_{0.85}$Ga$_{0.15}$As cladding layer. The layer-stack of the top laser waveguide consists of n-doped InGaP-contact layer, n-Al$_{0.61}$Ga$_{0.39}$As waveguide layer, an InGaAsP active region, p-Al$_{0.61}$Ga$_{0.39}$As waveguide layer, p-Al$_{0.85}$Ga$_{0.15}$As cladding layer and a p-GaAs cap. 

Multiple lateral tapers allow a guided mode to be coupled between the upper laser waveguide and the bottom BRW. Consequently, the device can then be operated as a laser, thereby providing pump light for the second-order nonlinear process. The necessary optical feedback for lasing operation can be provided by e.g. a distributed Bragg reflector (DBR) configuration as is shown in Ref. \cite{tenzler2025tunable}. In terms of coupling, adiabatic tapers enable low-loss coupling of a mode in one waveguide into a mode of another as long as the taper angle is sufficiently small avoiding conversion into unwanted, typically higher order, modes.
The coupling efficiency between two different waveguide modes follows the expression \cite{xia2005photonic}
\begin{equation}
	T_{ij}\propto \int dz\, A(z) e^{i (k_i - k_j)z}
	\label{Coupling_efficiency}
\end{equation}
with the coupling coefficient $A(z)$ depending on the modal overlap between the waveguide modes $i$ and $j$ and the propagation constants $k_i = 2\pi n_{\mathrm{eff}, i}/\lambda_0$ and $k_j = 2\pi n_{\mathrm{eff}, j}/\lambda_0$ determined by the effective indices $n_{\mathrm{eff}, i}$ and $n_{\mathrm{eff}, j}$. Therefore, efficient coupling requires matching of the effective indices of the modes, in addition to a nonvanishing modal overlap. 
\label{Section_Design}
\begin{figure}
	\centering
	\includegraphics[scale=0.6]{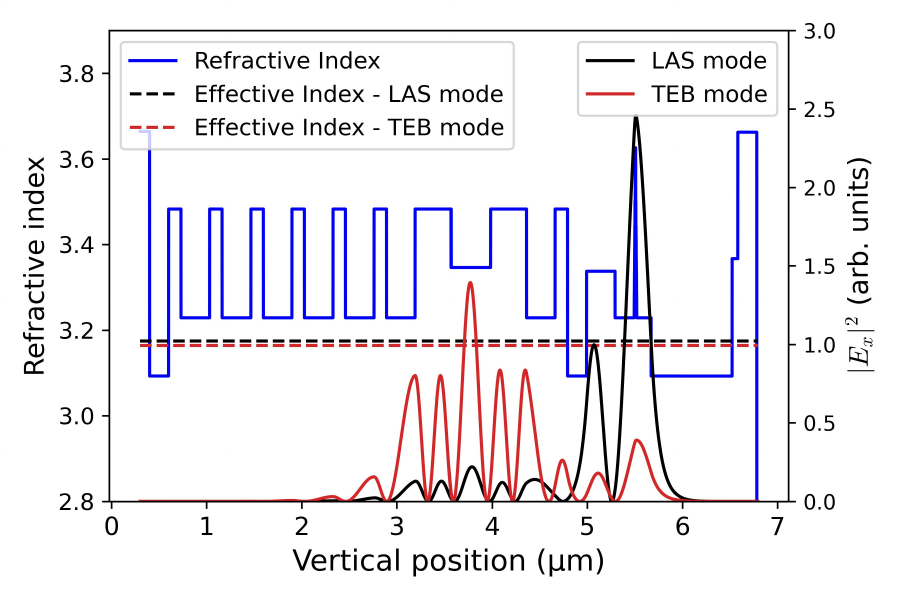}
	\caption{Vertical profiles of the refractive index (blue) and intensities of the LAS (black solid) and TEB (red solid) modes. The corresponding effective indices of the two modes are shown by the dashed lines.}
	\label{Vertical_profile}
\end{figure}

The vertical structure supports multiple modes at 775 nm. The most relevant are the fundamental TE-polarized lasing mode (LAS mode) in the upper laser waveguide and the TE-polarized Bragg mode (TEB mode) in the bottom BRW. The intensity profiles of both modes obtained from 1D simulations are shown in Fig.~\ref{Vertical_profile}. The effective index of the LAS mode ($n_{\mathrm{eff, LAS}}=3.173$) is slightly higher than that of the TEB mode ($n_{\mathrm{eff, TEB}}=3.165$). Thus, by tapering the upper laser waveguide, which reduces the effective index of the LAS mode, the effective indices of the LAS and TEB modes can be matched. It should also be noted that the vertical structure supports several guided modes, such as the fundamental mode (FM mode) of the bottom BRW and a parasitic mode (PAR mode) that is guided in the reflector layers of the bottom BRW. These two modes have significantly higher effective indices with $n_{\mathrm{eff, FM}}=3.428$ and $n_{\mathrm{eff, PAR}}=3.313$ than that of the TEB mode, which means that they can not be phase matched with fundamental modes around 1550 nm. As the effective indices of these parasitic modes are significantly higher than that of the LAS mode, no parasitic coupling with the LAS mode is expected.

In the context of the modal overlap, which is necessary for coupling between the waveguide modes according to Eq. (\ref{Coupling_efficiency}), it is important to note that the vertical structure of the bottom BRW was designed asymmetrically with six pairs below the core region and only one pair above it. Thus, the low thickness of the 'cladding' layers between the active region and the core of the BRW results in leakage of the TEB and the LAS modes into the corresponding region of the waveguide, therefore ensuring modal overlap between both waveguide modes. 

Despite this asymmetric design of the BRW, the waveguide supports the TEB mode at 775 nm, which can be phase matched to the TE-/TM-polarized fundamental modes at 1550 nm. On the other hand, high overlap of the LAS mode with the active region is important for laser performance. Therefore, an Al$_{0.85}$Ga$_{0.15}$As cladding layer is introduced below the n-contact, which increases the overlap of the lasing light with the active region. 

For the sake of completeness, we would like to note that, if the BRW is designed symmetrically with multiple reflector pairs above and below the core region of the BRW, then no coupling between the LAS and TEB mode can be observed. This is a result of the working principle of the Bragg reflector layers that destructively interfere light which is transmitted through them. As the effective indices of the LAS mode and the TEB mode are fairly similar, the modal overlap between the LAS and TEB modes vanishes.

It should be noted that the n-contact layer in between the BRW and the laser waveguide consists of lattice matched n-InGaP, analogous to Ref. \cite{koester2025design} which is transparent at 775 nm. This material was introduced due to the fact that all layers between the laser active region and the BRW core region must be transparent at 775 nm, as light must traverse through all layers during taper-induced coupling. Otherwise, the use of non-transparent layers would result in substantial absorption losses which would be detrimental for the device performance. In contrast, GaAs is not suitable as the n-contact material due to the low band-gap energy $E_{g,\mathrm{GaAs}}=1.42\,\mathrm{eV} < E_{\lambda=\mathrm{775 nm}}= 1.6\,\mathrm{eV}$ \cite{piprek2003semiconductor}. 

In the following section lateral tapers are introduced and the coupling efficiency between the LAS and TEB waveguide modes is investigated.

\section{Coupling efficiency of the lateral taper}
 
\begin{table}[htbp]
	\caption{Widths $W_{\mathrm{i}}$ and longitudinal extend $z$ of the lateral tapers and etch depths $d_\mathrm{i}$ in each taper segment. }
	\begin{tabular}{cccccc}
		\hline
		Taper segment&$z$ (µm)&$W_{\mathrm{L}}$ (µm)&$W_{\mathrm{LS}}$		(µm)&$W_{\mathrm{C}}$ (µm)&$W_{\mathrm{BRW}}$ (µm)\\
		\hline
		Laser ridge&0-30&3.0$\to$0.5&7.0$\to$4.5&11.0$\to$8.5&15.0$\to$12.5\\
		Laser waveguide&30-180&-&4.5$\to$0.5&8.5$\to$4.5&12.5$\to$8.5\\
		Contact layer&180-200&-&-&4.5$\to$0.5&8.5$\to$7.0\\
		BRW&200-220&-&-&-&7.0$\to$5.0\\
		Straight BRW&$>$220&-&-&-&5.0\\
		\hline
		&$d_{\mathrm{L}}$&$d_{\mathrm{LS}}$&$d_{\mathrm{C}}$&$d_{\mathrm{BRW,s}}$&$d_{\mathrm{BRW}}$\\
		\hline
		Etch depths (µm)&0.960&1.489&1.789&1.989&3.591\\
		\hline
	\end{tabular}
	\label{Taper_params}
\end{table}
~
\begin{figure}[t]
	\includegraphics[scale = 0.42]{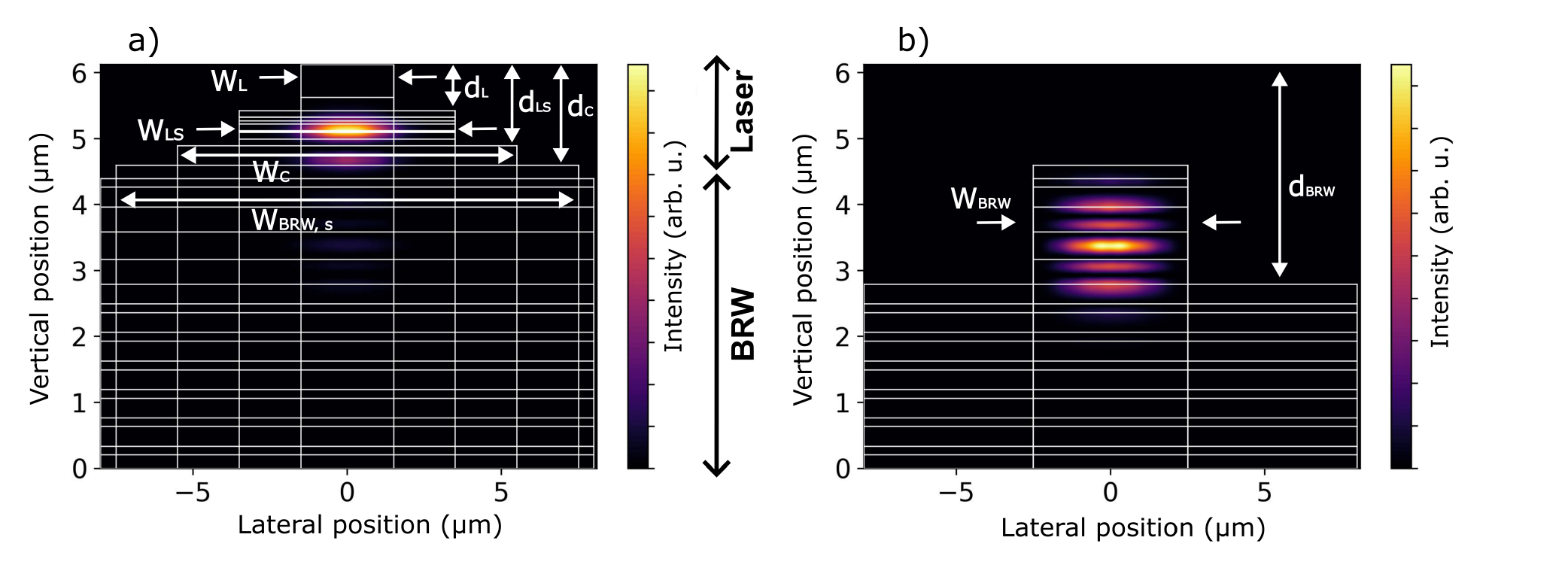}
	\caption{a) Intensity distribution of the LAS mode at the at the rear end of the taper at longitudinal position $z=0$. b)~Intensity distribution of the TEB mode at the front end of the taper at longitudinal position $z\geq 220$~µm. The etch depths are denoted by $d$ and the lateral waveguide widths are denoted by $W$.}
	\label{Mode plots}
\end{figure}
 
\begin{figure}[t]
	\includegraphics[scale = 0.44]{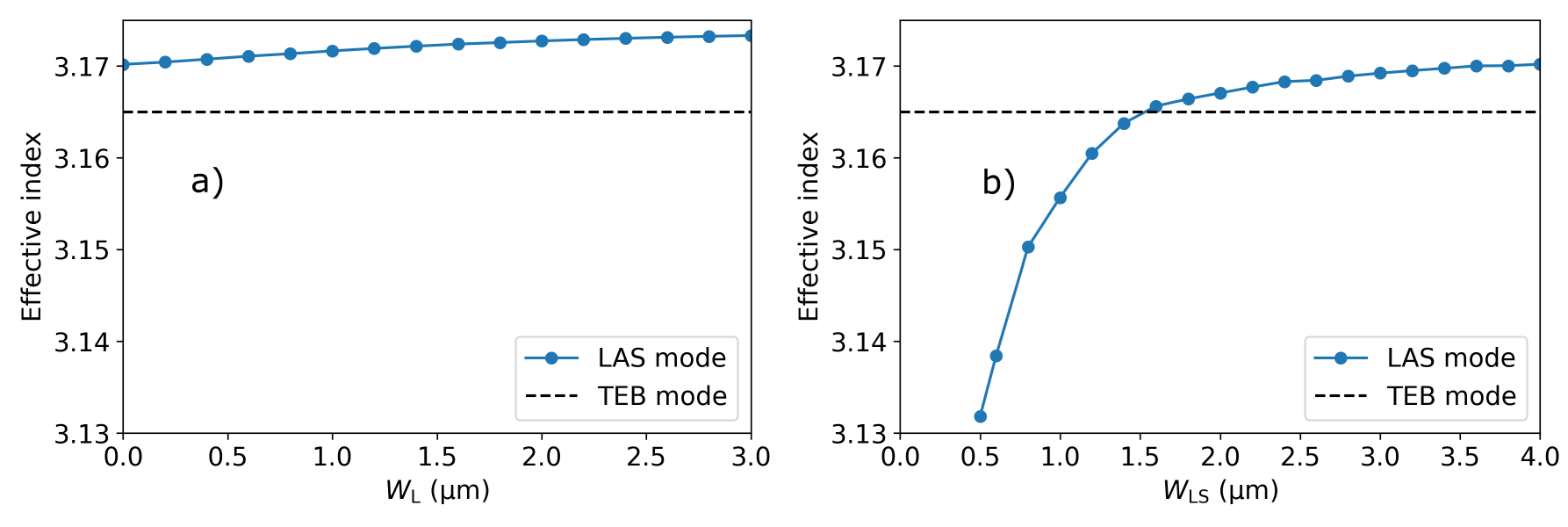}
	\caption{Effective index of the LAS mode (blue) for a) varying $W_{\mathrm{L}}$ analogous to the tapered region shown in Fig. \ref{Taper_plots}a between 0 µm and 30 µm and b) for varying $W_{\mathrm{LS}}$ analogous to the tapered region shown in Fig. \ref{Taper_plots}a between 30 µm and 180 µm. The dashed line represents the effective index of the TEB mode. The line connecting the data points has been added to provide a visual guide to the eye.}
	\label{Effective_indices}
\end{figure}
\label{Section_Simulation}
By performing 2D+$z$ simulations using the eigenmode expansion algorithm implemented in the software FIMMWAVE/FIMMPROP \cite{FIMMWAVE}, the coupling efficiency of the taper between the LAS mode and the TEB mode can be determined. The algorithm expands the field within a transverse cross section by a linear combination of the local waveguide modes. In longitudinal direction the taper is partitioned into several, small and longitudinally uniform cross sections, where the resulting transition conditions of the electric and magnetic fields at the interfaces of two adjacent cross sections yield transmission and reflection matrices. Combined with matrices describing the propagation in longitudinal direction, the total scattering matrix of the taper is calculated. The performance of a taper of length $L$ is determined by the transmission matrix element $|T^{0\to L}_{\mathrm{LAS, TEB}}|^2$ that represents the fraction of power that is coupled from the LAS mode into the TEB mode between the rear end of the taper ($z=0$) and the front end of the taper ($z=L$). 

The vectorial waveguide modes at each cross section are determined using the complex finite difference method (FDM) solver of FIMMWAVE. In order to speed up the simulation we make use of the lateral symmetry of the structure, as this halves the size of the simulation domain. In order to obtain results in a reasonable time frame, we introduce closed boundary conditions, specifically metal walls on top and bottom of the structure and magnetic walls on the sidewalls of the simulation domain. Furthermore, in order to avoid parasitic reflections at the computational boundaries, perfectly matched layers (PML) with a thickness of 0.2 µm are introduced at the sidewalls and the bottom of the simulation domain. The spatial resolution of the mesh is 0.1~µm in lateral direction and 0.01 µm in vertical direction. For these simulation parameters, the resulting distribution of the LAS mode at $z=0$ is shown in Fig. \ref{Mode plots}a and for the TEB mode in the BRW-RW is shown in Fig. \ref{Mode plots}b.

Before performing 2D+z simulations of the taper itself, regions of interest where coupling between the LAS mode and the TEB mode is expected can be identified based on the effective indices of the modes by varying the ridge widths of the structure. Fig. \ref{Effective_indices}a shows that the decrease of the width of the laser ridge ($W_{\mathrm{L}}$) has a negligible impact on the effective index of the LAS mode. As long as the effective index of the LAS mode remains above the effective index of the TEB mode, coupling in this region is expected to be negligible. On the other hand, a decrease in the width of the laser waveguide ($W_{\mathrm{LS}}$) shifts the effective index of the LAS mode below that of the TEB mode for $W_{\mathrm{LS}}<1.5\,$µm, see Fig. \ref{Effective_indices}b. Thus, tapering of the laser waveguide enables coupling between both modes.
\begin{figure}
	\includegraphics[scale = 0.44]{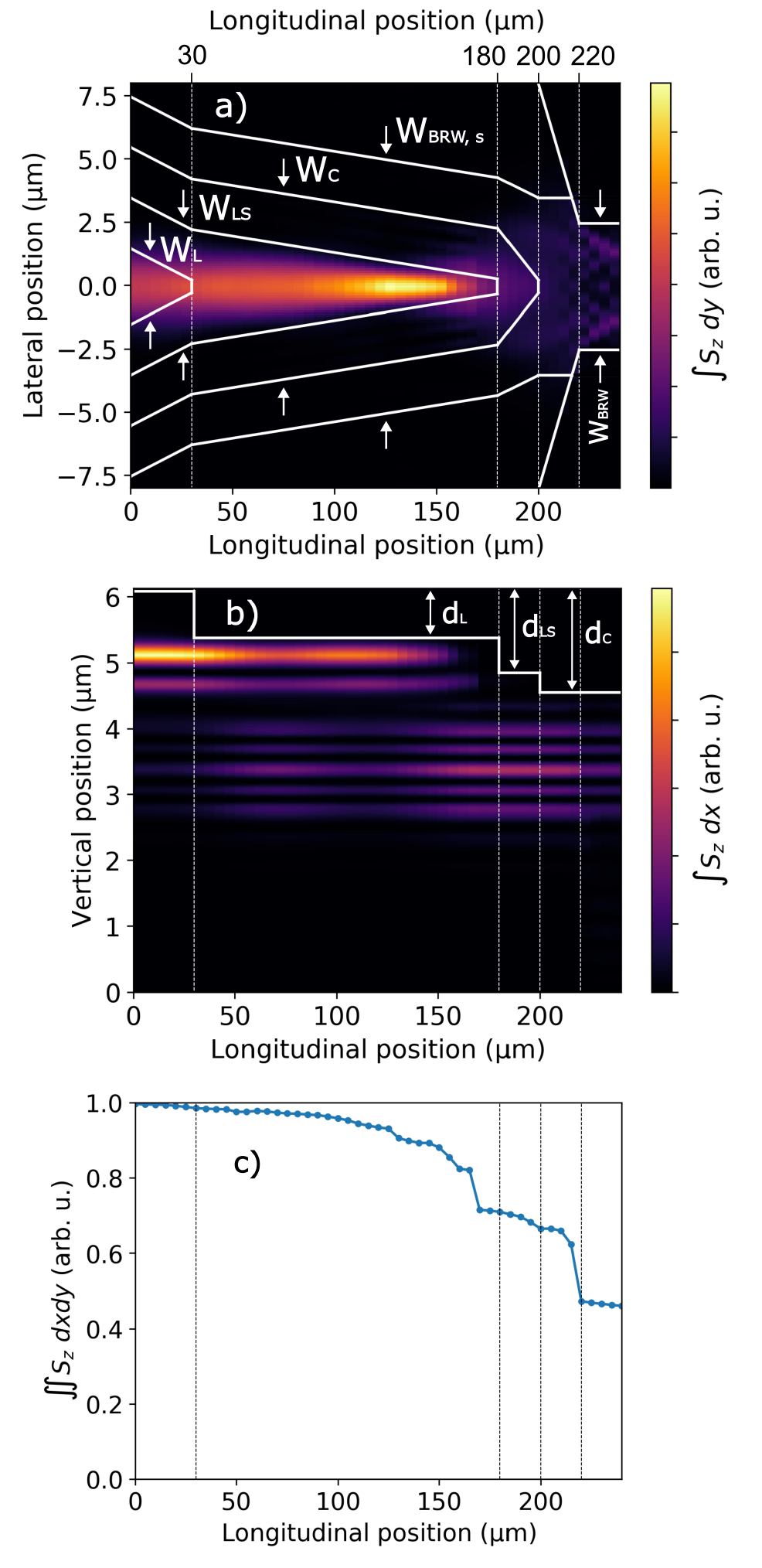}
	\centering
	\caption{a) Top-view of the structure. The color plot represents the vertically integrated longitudinal component of the Poynting vector. b) Side-view of the structure. The color plot represents the laterally integrated longitudinal component of the Poynting vector. c) Laterally and vertically integrated longitudinal component of the Poynting vector as a function of the longitudinal position. The solid horizontal and vertical line in a)-c) separate regions with different refractive indices and etch depths $d$, respectively. The dashed lines mark the edges of each taper segment according to Table \ref{Taper_params}. The lateral waveguide widths are denoted by $W$. }
	\label{Taper_plots}
\end{figure}

In order to realize efficient coupling between the LAS and TEB modes, multiple lateral tapers are introduced each with a different etch depth. The ridge width and length of each of these taper segments are shown in Table \ref{Taper_params}. In order to illustrate the power distribution within the structure, the color plots of the integral of the longitudinal component of the Poynting vector in the lateral direction are presented in Fig. \ref{Taper_plots}a and in the vertical direction in Fig. \ref{Taper_plots}b. The longitudinal component of the Poynting vector is defined as
\begin{equation}
	S_z := \frac{1}{2}\Re(E_x H_y^* - E_y^* H_x).
	\label{Poynting_vector}
\end{equation}
This quantity represents the energy flux in longitudinal direction, and thus the color plots highlight the regions where light is coupled between the upper and bottom waveguide modes throughout the taper.

In the first taper segment the p-contact layer and ridge of laser waveguide are removed. As the removal of the laser ridge only marginally reduces the effective index of the LAS mode, as shown in Fig. \ref{Effective_indices}a, the coupling in this region is negligible. Therefore, a 30 µm long taper segment is selected.

The second segment of the taper, in which the laser waveguide is removed, is primarily responsible for the coupling process, due to the effective index of the LAS mode being significantly decreased, see Fig. \ref{Effective_indices}b. As shown in Fig. \ref{Taper_plots}a, the coupling of light propagating in the LAS mode into waveguide modes of the bottom BRW waveguide can be observed, particularly towards the tip of the lateral taper at around 180~µm. In order to approach adiabatic power transfer, a length of 150~µm is selected for this taper segment. Subsequently, a third taper is used to remove the highly doped n-contact InGaP layer. This is due to the fact that the high refractive index of InGaP results in a significant fraction of power being guided in this layer, which would increase losses. 

It should be noted that the waveguide is not etched immediately to its optimal depth for the nonlinear process. Instead, an additional taper is introduced, that etches the cladding underneath the n-contact InGaP layer, in order to suppress coupling into high order lateral modes. 
In the final taper segment the BRW-RW is etched towards the optimal ridge width and etch depth required for the nonlinear process. As the coupling between modes is not of interest in these final two segments, the length of these taper segments is short and has been chosen to be 20 µm. Furthermore, in order to more realistically model the fabrication restrictions regarding the lithographic resolution of real devices, we consider the width of the tip of each taper to be non-zero (0.5 µm width) in our simulation. 

For this structure, we find a coupling efficiency between the LAS mode and the TEB mode of approximately $|T^{\mathrm{0\to L}}_{\mathrm{LAS,TEB}}|^2=28\%$. Furthermore, the coupling efficiency in the opposite direction is identical, i.e. $|T^{\mathrm{L\to 0}}_{\mathrm{TEB, LAS}}|^2=28\%$. Despite the introduction of the shallow etch step, roughly one fourth of the power is coupled into high order lateral waveguide modes which decreases the efficiency. This is shown by Fig. \ref{Taper_plots}c where the total power as a function of the longitudinal position remains at 50\% after the final taper segment. 

In the following section the photon pair generation of the BRW-RW through the SPDC process is investigated. 
\section{Properties of the BRW-RW}
\label{Section_Results}
\begin{figure}
	\centering
	\includegraphics[scale = 0.4]{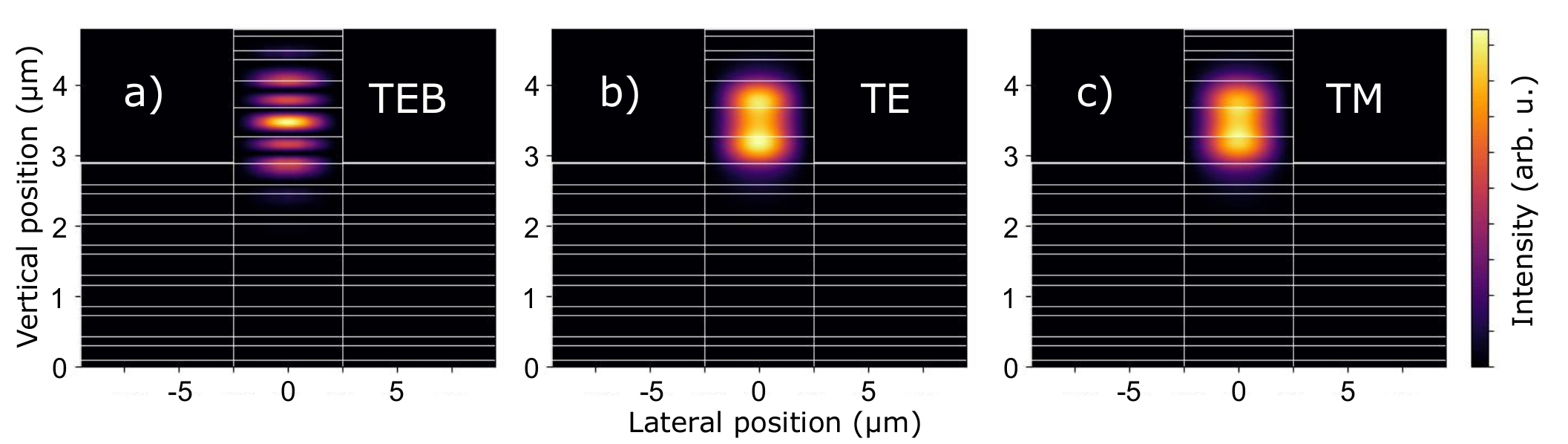}
	\caption{Color plots of the intensity distributions of the a) TEB mode at 775 nm. b) TE mode at 1550 nm. c) TM mode at 1550 nm in the BRW-RW for $z>220$ µm.}
	\label{BRW-RW_mode_profiles}
\end{figure}
\begin{figure}
	\centering
	\includegraphics[scale = 0.45]{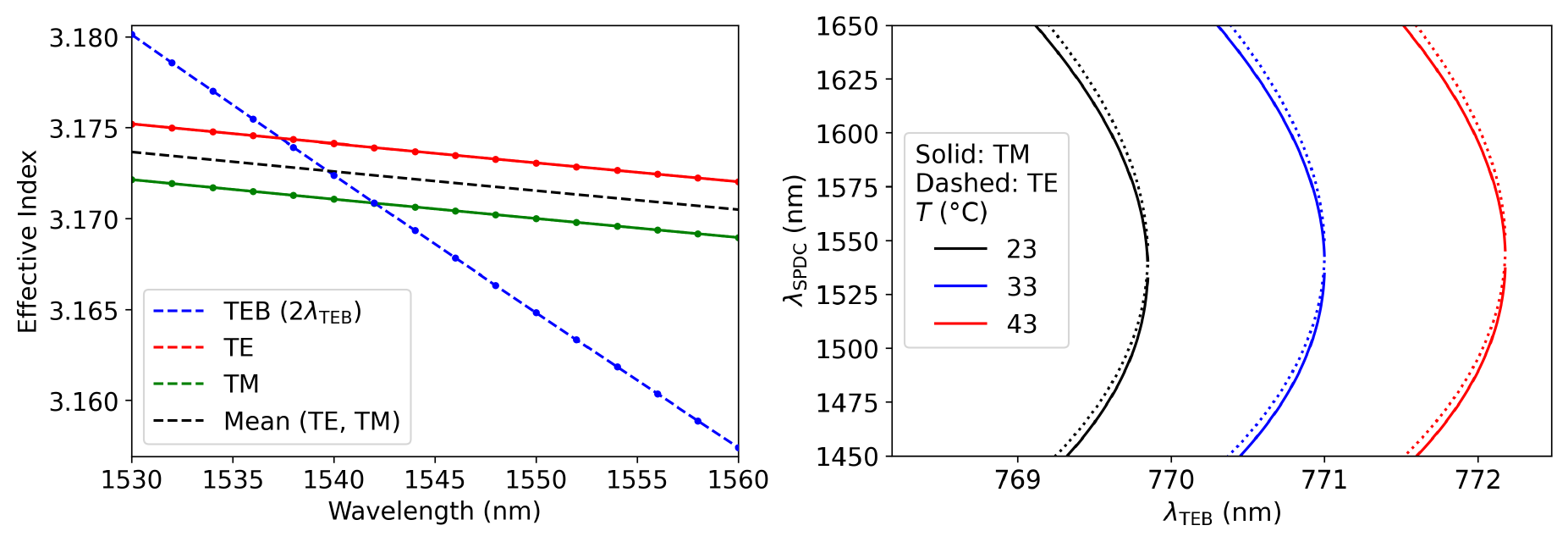}
	\caption{a) Effective index dispersion of the TEB mode (blue), the TE mode (red) and the TM mode (green) at $T=23\,\mathrm{^\circ C}$ in the BRW-RW. b) Phase diagrams of second order nonlinear processes for varying temperatures of the BRW-RW. }
	\label{Phase_diagram}
\end{figure}
\begin{figure}
	\centering
	\includegraphics[scale = 0.5]{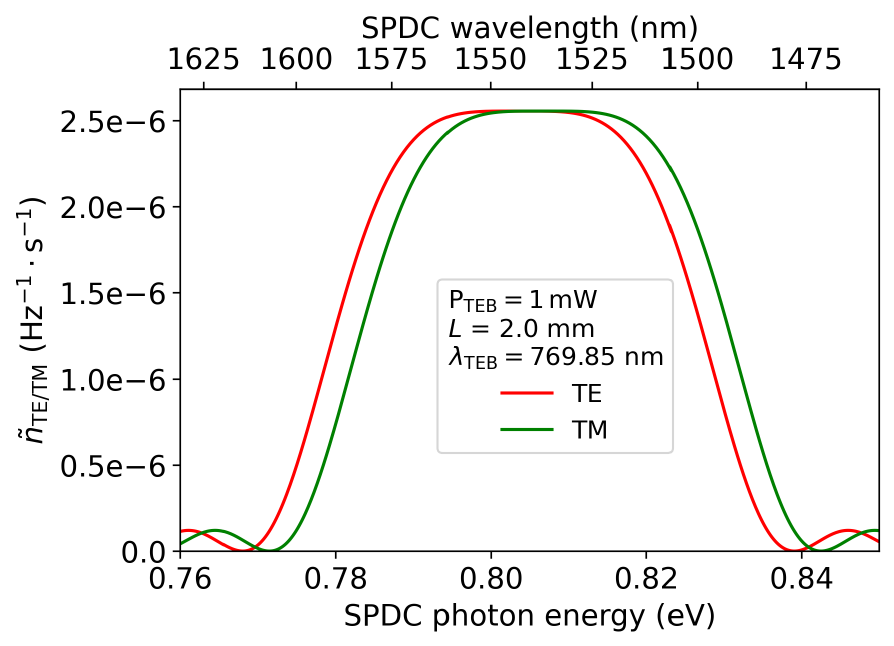}
	\caption{Spectral photon rate $\tilde{n}_{\mathrm{TE/TM}}$ of the frequency degenerate SPDC process for TE-/TM-polarized light for a 2 mm long BRW-RW assuming 1 mW power in the Bragg mode.}
	\label{Photon_pair_rate}
\end{figure}
After tapering, the bottom BRW-RW features a ridge width of 5 µm and is etched through the second matching layer. The waveguide supports the TEB mode at 775 nm and the fundamental modes at 1550 nm with the intensity profiles being shown in Fig. \ref{BRW-RW_mode_profiles}. 

The generation of photon pairs through the process of SPDC depends on the process being phase-matched. This is determined by the dispersion of the effective indices of the TEB mode around 775 nm and the fundamental modes around 1550 nm, which are shown in Fig. \ref{Phase_diagram}a). The corresponding phase diagram represents the solutions of the phase matching equation
\begin{equation}
	n_{\mathrm{eff,TEB}}(\omega_{\mathrm{TEB}})\omega_{\mathrm{TEB}} = n_{\mathrm{eff,TE}}(\omega_{\mathrm{TE}})\omega_{\mathrm{TE}} + n_{\mathrm{eff,TM}}(\omega_{\mathrm{TM}})\omega_{\mathrm{TM}}
	\label{Phase_matching}
\end{equation}
and is shown in Fig. \ref{Phase_diagram}b). The effective indices are determined by the 2D full vectorial FDM eigenmode solver implemented in FIMMWAVE \cite{FIMMWAVE}, while the refractive index model, its dispersion as well as the calculation of the nonlinear properties are implemented by a Python script based on models shown in Ref. \cite{tenzler2025theoretical}.  
The rate at which photon pairs are generated is determined by the modal overlap and the effective nonlinearity as shown in Ref. \cite{tenzler2025theoretical, abolghasem2011monolithic}. For our structure we find, $d_{\mathrm{eff}}=52\,\mathrm{pm/V}$ and $\xi=0.06$\,µ$\mathrm{m^{-1}}$. Thus, we estimate that a photon pair rate of $\dot{N}_{\mathrm{tot}}\approx1.7\cdot 10^{8}/\mathrm{s}$ is generated assuming an input power of 1~mW in a 2~mm long BRW-RW for frequency degenerate SPDC ($\lambda_{\mathrm{TEB}}\approx 770\,\mathrm{nm}$ @23\,°C). This rate is comparable with other passive BRW-RW as shown in Ref. \cite{horn2012monolithic}.

Additional tuning of the wavelength of phase matching is accomplished through temperature variation by $\approx 1.2\,\mathrm{nm/K}$, as shown in Fig \ref{Phase_diagram}b.

The photon pair rate $\tilde{n}_{\mathrm{TE/TM}}$ \cite{tenzler2025theoretical} for frequency degenerate SPDC is shown in Fig. \ref{Photon_pair_rate}. At twice the pump wavelength $\lambda_{\mathrm{TEB}}$, the rate at which TE-TM and TM-TE photon pairs are generated remains constant over a spectral range of approximately 30 nm. Beyond this regime, the rates deviate for the TE- and TM-polarized photons ($\tilde{n}_{\mathrm{TE}}\neq \tilde{n}_{\mathrm{TM}}$). The reason is a mismatch of the group velocity of the fundamental modes as shown in Refs. \cite{tenzler2025theoretical, zhukovsky2012generation}. Thus, when utilizing this device as a polarization entangled photon pair source, the optimal operating range is between 1525 nm and 1555 nm where the majority of photon pairs are produced and the rates are equivalent for both polarization states. 

\section{Conclusion}
\label{Section_Conclusion}
We designed and investigated a stacked waveguide structure consisting of a passive BRW at the bottom and an active laser waveguide on top by means of numerical simulations. In this active-passive structural approach the laser waveguide is doped, while the BRW remains undoped. The generation of light in the laser waveguide is achieved through injection of holes from a top p-contact and the injection of electrons from a lateral n-contact. The power of the lasing mode is transferred to the TE-polarized Bragg mode of the BRW-RW with the help of several lateral tapers providing pump light for the phase-matched, second-order nonlinear SPDC process. For the investigated structure we find a coupling efficiency between the LAS mode and the TEB mode of $|T^{\mathrm{0\to L}}_{\mathrm{LAS,TEB}}|^2=28\%$. 

The stacked waveguide approach has several advantages. For instance SPDC is achieved without external optical pumping. Additionally, losses from free carrier absorption of the photon pairs and parasitic luminescence are avoided due to the passive BRW-RW being undoped and electrically unbiased. The photon pair emission rate generated by frequency degenerate SPDC was determined based on calculated properties of the BRW-RW and reads $\dot{N}_{\mathrm{tot}}\approx1.7\cdot 10^{8}/\mathrm{s}$ for a 2 mm long BRW-RW assuming 1 mW power in the TEB mode. The photon pair rate of the stacked waveguide is therefore comparable to other passive BRW-RW operating at a similar wavelength.

\section*{Acknowledgment}
We gratefully acknowledge support from the German Federal Ministry of Research, Technology and Space (BMFTR) grant 16KIS1768 and 16KIS1765K as part of VOMBAT. 

\bibliography{Literature}

\end{document}